\documentclass[12pt]{iopart}
\usepackage{iopams}  

\usepackage{epsfig,graphicx,float,pst-all, rotating}
\usepackage{hyperref}
\usepackage{amssymb}
\usepackage{multirow}
\usepackage[english]{babel}
\usepackage{comment}

\usepackage[numbers,square,sort&compress]{natbib}
\usepackage{color}
\usepackage{xcolor}
\usepackage{soul}

\newcommand{\be}{\begin{equation}} 
\newcommand{\ee}{\end{equation}}
\newcommand{\bea}{\begin{eqnarray}} 
\newcommand{\eea}{\end{eqnarray}}
\newcommand{\bc}{\begin{center}}
\newcommand{\ec}{\end{center}}

\newcommand{\eps}{\varepsilon}

\begin{document}

\title{Continuum discretized BCS approach for weakly bound nuclei} 

\author{J. A. Lay}

\address{ Dipartimento di Fisica e Astronomia ``Galileo Galilei", Universit\`{a} di Padova, via Marzolo, 8, I-35131 Padova, Italy}
\address{INFN, Sezione di Padova, via Marzolo, 8, I-35131 Padova, Italy}
\ead{lay@pd.infn.it}

\author{C. E. Alonso}
\address{ Departamento de F\'isica At\'omica, Molecular y Nuclear, Facultad de F\'isica, Universidad de Sevilla, Apartado 1065, 41080 Sevilla, Spain}

\author{L. Fortunato and A. Vitturi}
\address{ Dipartimento di Fisica e Astronomia ``Galileo Galilei", Universit\`{a} di Padova, via Marzolo, 8, I-35131 Padova, Italy}
\address{INFN, Sezione di Padova, via Marzolo, 8, I-35131 Padova, Italy}


\vspace{10pt}
\begin{indented}
\item[]\today
\end{indented}

\begin{abstract}

The Bardeen-Cooper-Schrieffer (BCS) formalism is extended by including the single-particle continuum in order to analyse the evolution of pairing in an isotopic chain from stability up to the drip line. We propose a continuum discretized generalized BCS based on single-particle pseudostates (PS). These PS are generated from the diagonalization of the  single-particle Hamiltonian within a Transformed Harmonic Oscillator (THO) basis. The consistency of the results versus the size of the basis is studied. The method is applied to neutron rich Oxygen and Carbon isotopes and compared with similar previous works and available experimental data. 
We make use of the flexibility of the proposed model in order to study the evolution of the occupation of the low-energy continuum when the system becomes weakly bound.  We find a larger influence of the non-resonant continuum as long as the Fermi level approaches zero.

\end{abstract}

%
\vspace{2pc}

%
\submitto{\JPG}
%
%
%

\section{Introduction}

Nuclei far from the $\beta$-stability line are currently one of the most active fields in nuclear physics. New radioisotope ion beam facilities are nowadays producing short-lived nuclei at intermediate masses reaching the neutron drip-line, thus venturing into a hitherto unknown domain of nuclear phenomena. 
In this region of the nuclear chart, the ground state energy and the Fermi level approach the threshold for neutron emission, therefore, the proper inclusion of continuum states becomes progressively more and more important. 
The treatment of the continuum represents a challenge on both theoretical and computational grounds. 
Different representations of the continuum have been used and compared with few-body models \cite{Das09,Del07,Mor06,Mor09,Lay10,Dru10,For14,Rod08,Cas13}. These models are accurate in nuclei with a strong clusterization into few stable fragments as it occurs in the light region of the Segr\`e Chart. When moving to the study of heavier isotopes, these models require some extension or additional ingredients~\cite{deD14,Mor12a,Lay14,Del13,Cap14,Pot14,Des13}. 

On the other hand, the intermediate mass region of the Segr\`e Chart is the main region of application of mean field approaches and the study of pairing. However, known isotopes in this region are strongly bound so that traditional formulations of many-body models do not include the continuum. In the field of many-body systems the first attempt to include the single particle continuum is the Continuum Shell Model (CSM) \cite{CSM,Esb97,Vol06}. From this CSM, a BCS formulation was presented in~\cite{San97} and different variations have been explored either using the complex energy plane~\cite{IdB06,Dus07}, diagonalizing in a box~\cite{IdB08}, or making use of continuum single particle level densities (CSPLD)~\cite{IdB12,IdB12b}. Other formulations, which have been generalized to include the continuum or at least its resonant part, are Hartree-Fock-BCS (HF-BCS)~\cite{San00}, Hartree-Fock-Bogoliubov (HFB)~\cite{Zha11,Gra01,Ham03,Ham04}.  Another approach is the complex scaling method, in which a unitary transformation of the cut in the complex energy plane eases the treatment of resonances of the many-body system~\cite{Myo07,Myo12}. These resonances can also be studied with the Gamow Shell Model (GSM)~\cite{Mic02} together with its coupled-channel representation (GSM-CC)~\cite{Jag14,Fos15}. Finally also configuration-space Monte Carlo method (CSMC) has been recently extended to the use of pairing within the continuum~\cite{Lin15}.

In some of these formulations, there is a separate treatment of resonant and non-resonant parts of the continuum. Also regarding the discretization method for the continuum, only box boundary condition or the Berggren representation in the complex energy plane are predominantly used. 
Since one of the main problems in these calculations is the computational cost, looking at a number of different basis sets in order to optimize the convergence is a very important factor. It can be also interesting to check the stability of the many-body approach versus different ways of discretizing the continuum.

Therefore, we propose in this work the use of the Transformed Harmonic Oscillator (THO) basis for the discretization of the continuum into a Generalized BCS formalism. The prescription used here will be that of Karataglidis \textit{et al.}~\cite{Amos} rather than 
 that of the original work by Stoitsov \textit{et al.}~\cite{Sto98a}. This THO basis has been shown to reduce the computational cost in nuclear reactions within three- and four-body formalisms~\cite{Lay12,deD14,Rod08,Cas13}. This,  together with the flexibility of the THO basis shown in~\cite{Amos,Lay10} is crucial here for exploring the role of the continuum when a nucleus becomes more and more weakly bound. Another important fact is the possibility of treating both resonant and non-resonant part of the continuum in a natural way and on the same footing. This makes calculations simpler and more straightforward. 

A HFB formalism can give a more accurate description, solving some of the problems of the BCS plus continuum~\cite{Dob96,Kru01,Dea03}. However, this choice implies an increase of the complexity of the calculation and, therefore, a more difficult analysis of the results. An alternative option is to keep the essence of the HFB equations in a simplified model as done by I. Hamamoto and B. R. Mottelson~\cite{Ham04,Ham03}.

The present work is structured as follows. The method is described in Sec.~\ref{meth}. First, we recall how to discretize the single-particle continuum with the THO basis in subsection~\ref{tho}. Later, we describe the Generalized BCS formalism using the THO pseudostates with a Density Dependent Delta Interaction in subsection~\ref{cgBCS}. Finally, we apply the formalism to Oxygen and Carbon 
isotopes in Sec.~\ref{res}.

\section{ \label{meth} Methodology}

\subsection{ \label{tho} Discretization of the single-particle continuum}

In this subsection, we briefly review the features of the pseudo-states (PS) basis used
in this work. The THO basis is meant to describe the states of a composite system consisting on two
interacting inert fragments, such as a valence particle
(proton/neutron) and a spherical and stable core. In this case,  the core+valence Hamiltonian
is simply given by: 
\be
h_{s.p.}= T_r + V_{vc}({r}) 
\label{hsp}
\ee
where $\vec{r}$ is the relative coordinate between the valence particle and the
core, $T_r$ the core-valence kinetic energy operator and  
$V_{vc}({r})$ is the interaction between the valence particle and the core. 
The eigenstates of this Hamiltonian can be characterized by the
excitation energy ($\varepsilon$) and the set of quantum numbers
$\{\ell, s, j, m \}$, which correspond to the orbital angular momentum
($\vec{\ell}$), the valence spin ($\vec{s}$) and their vector sum ($\vec{j}=\vec{\ell} +
\vec{s}$). For a central potential with, possibly, a spin-orbit term,
these states can be written as: 
\begin{equation}
\phi_{\varepsilon,\ell,s,j,m}(\vec{r}) = R_{\varepsilon,\ell, j}(r)  {\cal
  Y}_{\ell s j m}(\hat{r})   
\label{wfgen}
\end{equation}
where  ${\cal Y}_{\ell s j m}(\hat{r}) = \left[ Y_{\ell}(\hat{r})
  \otimes \chi_s \right]_{jm}$, with  $\chi_s$  a spin function. The
radial functions $R_{\varepsilon,\ell, j}(r)$ can be obtained by solving the
Schr\"odinger equation with appropriate boundary condition
for bound ($\varepsilon < 0$) or unbound ($\varepsilon > 0$)
states. Alternatively, these functions  can be obtained by
diagonalizing the Hamiltonian   (\ref{hsp})  in a discrete
basis. Since any complete basis  will be infinite, this procedure is
not feasible in practice unless the basis is  truncated. By doing so,
one obtains a finite (and approximate) expansion of the functions
$R(r)$ in the selected basis. If the basis  
functions are denoted by $\varphi_{n,\ell,s,j,m}(\vec{r})= R^{basis}_{n,\ell}(r) {\cal
  Y}_{\ell s j m}(\hat{r})$, we will have: 
\begin{equation}
R_\alpha (r)= \sum_{n=1}^{N}  c_{\alpha,n} R^{basis}_{n,\ell}(r) 
\end{equation}
where $\alpha\equiv \{\varepsilon,\ell,j\}$ and  $N$ is the number
of states retained in the basis.

As already mentioned, there are many possible choices for
the  basis functions $\{\varphi_n\}$ (Gaussian, harmonic oscillator,
Laguerre, etc). In this work we use the transformed
harmonic oscillator (THO) basis, obtained from the harmonic oscillator
basis with an appropriate local scale transformation (LST) as originally proposed by Stoitsov \textit{et al.}~\cite{SP88,PS91}.

If the LST function is  
denoted by $s(r)$, the THO states are obtained as 
\begin{equation}
\label{eq:tho}
R ^{THO}_{n, \ell}(r)= \sqrt{\frac{ds}{dr}} R^{HO} _{n, \ell}[s(r)],
\end{equation}
where $R ^{HO}_{n, \ell}(s)$ is the radial part of the usual HO functions. 
With the criterion given above, the LST is indeed not unique. Here, we adopted a parametric form for the LST from Karataglidis \textit{et al.}
\cite{Amos} 
\begin{equation}
\label{lstamos}
s(r)  = 
 \left[  \frac{1}{   \left(  \frac{1}{r}
    \right)^m  +  \left( 
\frac{1}{\gamma\sqrt{r}} \right)^m } \right]^{\frac{1}{m}}\ ,
\end{equation}
that depends on the parameters $m$ and $\gamma$. The extension of $R ^{HO}_{n, \ell}(s)$ will also depend on the oscillator length
$b$. Note that,  
asymptotically, the function  $s(r)$ behaves as 
$s(r)\sim \gamma \sqrt{r}$
and hence the functions obtained by applying this LST to the HO basis
behave at  
large distances as 
$\exp(-\gamma^2 r /(2 b^2))=\exp(-k_\mathrm{eff} r)$. Therefore, the ratio
$\gamma/b$ can be related to an effective linear momentum, 
$k_\mathrm{eff}=\gamma^2 / (2 b^2)$, which  
governs the asymptotic behavior of the THO functions. As the ratio
$\gamma/b$ increases, the radial extension of the basis decreases and,  
consequently, the eigenvalues obtained upon diagonalization of the
Hamiltonian in the THO basis tend to concentrate at higher excitation
energies. Therefore, $\gamma/b$ determines the density of eigenstates
as a function of the excitation energy. In all the calculations
presented in this work, the power $m$ has been taken as $m=4$. This
choice is discussed in  Ref.~\cite{Amos} where the authors found that
the results are weakly dependent on $m$.

Note that, by construction, the family of functions  
\( R ^{THO}_{n, \ell}(r) \) constitute a complete orthonormal set.
Moreover, they decay exponentially
at large distances, thus ensuring the correct asymptotic behaviour
for the bound wave functions. In practical calculations a finite set
of functions (\ref{eq:tho})  
is retained, and the single-particle Hamiltonian is
diagonalized in this truncated basis with $N$ states,  
giving rise to a set of  eigenvalues and their associated
eigenfunctions, denoted respectively by $\left\{\varepsilon_n \right\}$ and 
$\{\varphi^{(N)}_{n, \ell}(r)\}$ ($n=1,\ldots,N$). As the basis size
is increased,  the eigenstates with negative energy
will tend to the exact bound states of the system, while those
with positive eigenvalues can be regarded as a finite representation
of the unbound states.

Generalizations of this basis can be found for three-body systems~\cite{Rod08} and for two-body systems with core excitations~\cite{Lay12} with great success. Nevertheless, this is the first time this basis is applied to the study of pairing in the continuum.

\subsection{ \label{cgBCS} Generalized BCS with pseudostates}

Now we solve the many-body Hamiltonian of ${\cal N}$ neutrons moving in a mean field. In second quantization this Hamiltonian reads~\cite{RS}:  

\begin{equation}
{\cal H}=\sum h_{s.p.}+\sum_{klk'l'}v_{klk'l'}a^{\dagger}_{k}a^{\dagger}_{l}a_{k'}a_{l'} .
\end{equation}
In BCS the two-body interaction is reduced to those situations where two nucleons form a strongly correlated pair of time reversal states. Therefore, if we define the pair creation (and corresponding annihilation) operator:

\begin{equation}
P^{\dagger}_{j}=\sum_{m>0}(-1)^{j-m}a^{\dagger}_{jm}a^{\dagger}_{j-m},
\end{equation}
the previous Hamiltonian reads:
\begin{equation}
{\cal H}=\sum h_{s.p.}-\sum_{jj'}G_{jj'}P^{\dagger}_{j}P_{j'} ,
\end{equation}
where $G_{jj'}$ is a state-dependent coefficient that replaces $v_{klk'l'}$.

One should add to the sum over the different bound states, an integral over the different momentum states defined in the single-particle continuum . However, with the use of a discretized basis, we can turn this integral into a sum over different pseudostates. Notice that the pair operator is not uniquely defined with $j$. For each $j$ we would have $N$ pseudostates and, therefore, $N$ different pair operators defined by $\nu \rightarrow \{n,j\}$, where $n$ is the label ordering in energy the pseudostates. From these $N$ pair operators, we will have one operator with associated negative single particle energy per each bound state present in $h_{s.p.}$ and the rest up to $N$ will have an associated positive single particle energy so that they will create a Cooper pair in the continuum. The final number will be smaller than $N$ times the number of angular momenta considered since we have also imposed a maximum energy cut off.

Therefore, we can continue with the traditional definition of the BCS function:

\begin{equation}
|BCS\rangle=\Pi_{\nu}(u_{\nu}+v_{\nu}P^{\dagger}_{\nu})|0\rangle .
\end{equation}
Finally, we can obtain the corresponding BCS equations by minimizing the Hamiltonian ${\cal H}$ with constraint on the number of particles, i.e.:

\begin{equation}
\langle BCS | {\cal H} - \lambda {\cal N} |BCS \rangle ,
\end{equation}
where $\lambda$ is the Fermi level. The problem can be recast into solving the following set of non-linear coupled equations:

\begin{eqnarray}
\left\lbrace
\begin{array}{ccc}
\Delta_{\nu} & = & \frac{1}{4}\sum_{\nu'}\frac{(2j'+1)G_{\nu\nu'}\Delta_{\nu'}}{\sqrt{(\eps_{\nu'}-\lambda)^2+\Delta_{\nu'}^2}} \\ ~ \\
{\cal N} & = & \sum_{\nu}(2j+1)v_{\nu}^2 ,
\end{array}\right.
\end{eqnarray}
where $\eps_{\nu}$ is the single-particle energy and the coefficients $u$ and $v$ are given by:

\begin{eqnarray}
v_{\nu}^2 & = & \frac{1}{2}\left( 1-\frac{\eps_{\nu}-\lambda}{\sqrt{(\eps_{\nu}-\lambda)^2+\Delta_{\nu}^2}} \right) , \\
u_{\nu}^2 & = & \frac{1}{2}\left( 1+\frac{\eps_{\nu}-\lambda}{\sqrt{(\eps_{\nu}-\lambda)^2+\Delta_{\nu}^2}} \right) .
\end{eqnarray}

\subsection{The pairing interaction}

In order to calculate the values of $G_{\nu\nu'}$ we make use of a Density Dependent Delta Interaction (DDDI). For a general radial dependence of the delta interaction:

\begin{equation}
V_{res}(\vec{r}_{1}-\vec{r}_{2})=V(r)\delta (\vec{r}_{1}-\vec{r}_{2}) ,
\end{equation} 
the final values of $G_{\nu\nu'}$ will be:

\begin{equation}
G_{\nu\nu'}=\frac{-1}{2(4\pi)}\int_{0}^{\infty} |R_{\nu}(r)|^{2}|R_{\nu'}(r)|^{2} V(r) r^2 dr ,
\end{equation}
where $R_{\nu}(r)$ are radial single-particle wavefunctions. The minus sign ensures that the values of $G$ are defined positive for any attractive interaction. In our case, the radial form of the interaction is:
\begin{equation}
V(r)=\left[V_{0}-V_{I}\left(\frac{\rho_{0}}{1+\exp{\frac{r-R_{0}}{a_{0}}}}\right)^{\eta}\right] ,
\end{equation}
which can be accomodated to the more traditional form:
\begin{equation}
V(r)=V_{0}\left[1-\left(\frac{\rho(r)}{\rho_{0}}\right)^{\eta}\right] ,
\end{equation}
for a certain selection of the parameters.

We can compare these results with the use of a constant pairing $G$. In \cite{IdB08,IdB12,IdB12b}, it has already been shown that, in order to keep a constant $G$, a renormalization of the pairing strength in the continuum is needed. In that sense, the results from \cite{IdB08,IdB12,IdB12b} cannot be directly compared with the use of a constant $G$ here. With the use of a DDDI we hope to recover this renormalization in a natural way without distinguishing resonant and non-resonant continuum. This represents a major advantage of the present approach.

Finally, it should be said that the density $\rho(r)$ is taken from a phenomenological Woods-Saxon form adjusted to each case. No self-consistent recalculation of the density is considered. 

\section{\label{res} Results}

\subsection{Oxygen isotopes}

The isotopic chain of oxygen is the heaviest in atomic number with a well-known position of the neutron drip-line. There are two isotopes, $^{26}$O and $^{28}$O, with experimentally measured negative two-neutron separation energies (S$_{2n}$), meaning that both nuclei are unbound with respect to two-neutron emission~\cite{nndc}.

\begin{figure}
\hspace{0.15\textwidth}
\includegraphics[width=0.5\textwidth]{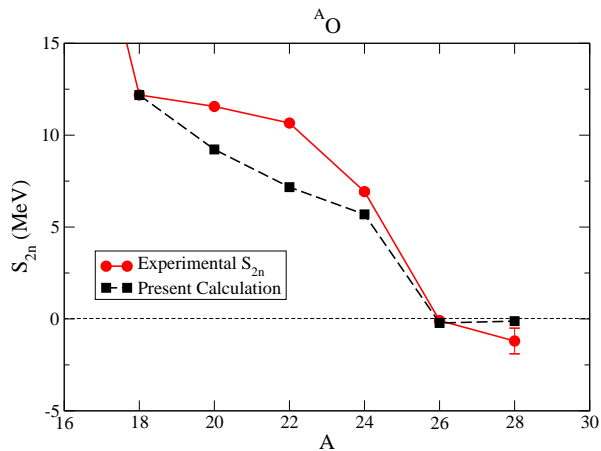}
\caption{\label{s2noxygen} (Color online) Two-neutron separation energies for the even neutron rich oxygen isotopes. Experimental points contain error bars (sometimes invisible at this scale).}
\end{figure}

In our calculations, we took a single-particle Hamiltonian able to reproduce the spectrum of $^{17}$O taking the $n$-$^{16}$O potential from Ref.~\cite{Spa00}. We will include all partial waves with $\ell\leq2$. For constraining the pairing interaction, we use a volume term with the geometry of this potential and a strength that reproduces the S$_{2n}$ in $^{18}$O. The strength of $V_{I}$ is later renormalized for the following even oxygen isotopes so that it follows the known $1/A$ dependence. The energy cut-off is set at 20~MeV. Therefore, all free parameters are tied to the knowledge of stable isotopes. We then add one, two, three, four, five and six pairs of neutrons to the $^{16}$O core  in order to reach $^{28}$O. The theoretical values obtained within the present framework for the two-neutron separation energies compared with the experimental ones taken from~\cite{nndc} are shown in Fig.~\ref{s2noxygen}. Recently, within the coupled-cluster theory~\cite{Jan14} it has been possible to look at the binding energies and the spectra of Oxygen (and also Carbon) isotopes. In our formalism we concentrate on binding energies focusing on the evolution of superfluidity and the role of continuum.

\begin{figure}
\hspace{0.15\textwidth}
\includegraphics[width=0.5\textwidth]{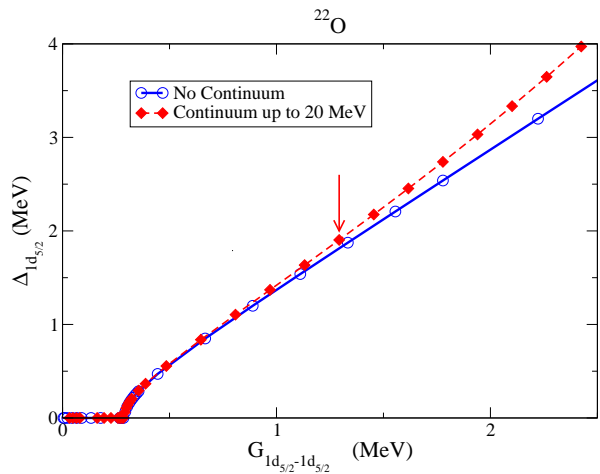}
\caption{\label{delta22o} (Color online) Value of $\Delta$ for the $1d_{5/2}$ orbit versus pairing strength for Cooper pairs in the same $1d_{5/2}$ orbit for $^{22}$O with (dashed line) and without (solid line) inclusion of the continuum. The red arrow show the pairing strength used in the present calculations for $^{22}$O.  }
\end{figure}

\begin{figure*}
\hspace{0.15\textwidth}
\includegraphics[width=0.5\textwidth]{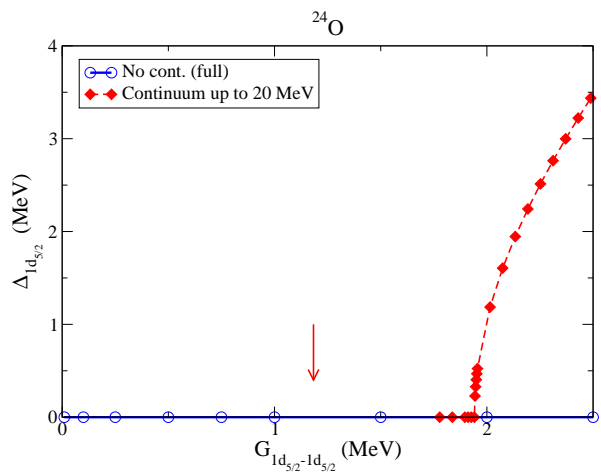}
\caption{\label{delta24o} (Color online) Value of $\Delta$ for the $1d_{5/2}$ orbit versus pairing strength for Cooper pairs in the same $1d_{5/2}$ orbit for $^{24}$O with (dashed line) and without (solid line) inclusion of the continuum.  The red arrow show the pairing strength used in the present calculations for $^{24}$O.}
\end{figure*}

In this calculation, the structure of $^{24}$O is not superfluid. Last bound orbitals $1d_{5/2}$ and $2s_{1/2}$ are fully occupied, so that the continuum and also the $d_{3/2}$ resonance are not significantly populated. Therefore we choose the closest isotope $^{22}$O where the unoccupied 2n-hole in the $1d_{5/2}$ or $2s_{1/2}$ orbits enhances the effect of  pairing, thus increasing the population of the continuum. This can be understood by looking at the $\Delta_{\nu}$ found for the present value of the pairing strength as shown in Fig.~\ref{delta22o} for $^{22}$O and in Fig.~\ref{delta24o} for $^{24}$O.

In these figures we plot $\Delta_{\nu}$ versus $G_{\nu\nu}$ of the single-particle ground state $1d_{5/2}$. The red arrow in these figures indicates the pairing  strength used in the present calculations for these nuclei. We see that in $^{22}$O the arrow indicates a point located after the transition to superfluidity, whereas in $^{24}$O the arrow is still far from the critical point. $^{24}$O appears, therefore, as a closed sub-shell nucleus. In~\cite{Lin15}, they applied the CSMC method also to $^{24}$O obtaining a really small impact of the continuum on the total binding energy in agreement with what is found here. We also compare with calculations without the continuum in both figures. Strong deviations from this no-continuum reference are not seen unless one forces the strength of the pairing well beyond its commonly accepted values. This is something to be expected, if we take into account that the continuum threshold is at $4.14$~MeV in $^{17}$O, and this is a gap to be overcome with the help of pairing.   One should keep in mind that the single-particle spectrum varies along the isotopic chain, so that by doing HF or similar calculations we expect this continuum threshold to be reduced. Therefore, there is still a chance for $^{24}$O to be superfluid or at least closer to superfluidity.


\begin{figure}
\hspace{0.15\textwidth}
\includegraphics[width=0.5\textwidth]{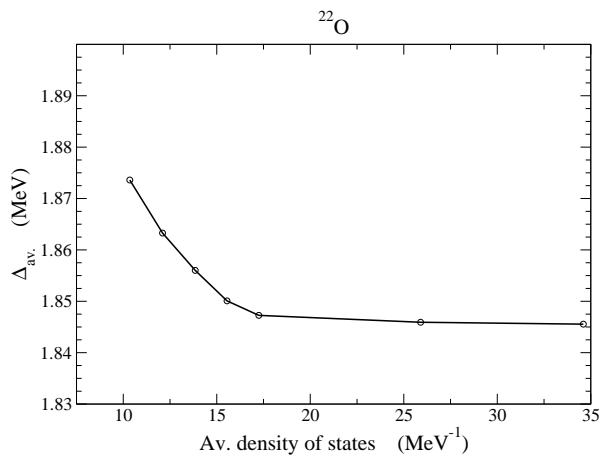}
\caption{\label{avdelta} Convergence of $\Delta_{\rm av.}$ with the size of the basis. The X axis represents the number of levels in the calculation divided by the cut off energy. The different points correspond to $N$=$\{60,70,80,90,100,150,200\}$. }
\end{figure}

In order to study the convergence properties, we show, for instance, the average $\Delta_{\rm av.}$ in $^{22}$O with respect to the size of the basis. This quantity is defined as the weighted sum over the corresponding occupations~\cite{Dob96}:

\begin{equation}
\Delta_{\rm av.}=\frac{\sum_{\nu} (2j+1) \Delta_{\nu}v^2_{\nu}}{\sum_{\nu} (2j+1) v^2_{\nu}} =\frac{\sum_{\nu} (2j+1) \Delta_{\nu}v^2_{\nu}}{{\cal N}}.
\end{equation}

Since $N$ is not a direct measurement of the final number of states considered in the calculation, we define an average density of states. For each partial wave we keep $N_{j}\leq N$ pseudostates, those with energies smaller than the energy cut off $E_{cut}$. Therefore, a way to measure an average density of states is:
\begin{equation}
{\rm Av.~~density~~of~~states } = \frac{\sum_{j} {N}_{j}}{E_{cut}} .
\end{equation}
The result is shown in Fig.~\ref{avdelta}.
One can see that the convergence is very good. These results have been obtained for the following parameters of the basis: $\gamma=2.1$~fm$^{1/2}$ and $b=2.1$~fm. For the sake of convenience, we use $N$=100 for all the remaining calculations in this work.

\begin{figure}
\hspace{0.15\textwidth}
\includegraphics[width=0.5\textwidth]{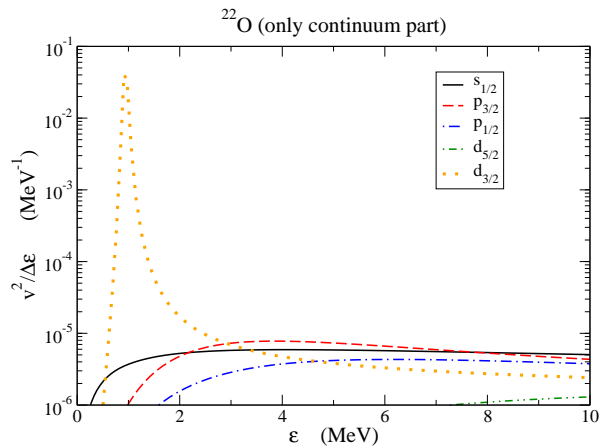}
\caption{\label{occ} (Color online) Density of occupation in the continuum for $^{22}$O for the different partial waves.}
\end{figure}

We show in Fig.~\ref{occ} the occupancies $v^{2}_{\nu}$ in the continuum for the different partial waves. We clearly see the dominant role of the low-lying $d_{3/2}$ resonance creating a peak around $1$~MeV with an occupation several orders of magnitude larger than those of the non-resonant continuum.  In a discrete set the direct sum of these values should give the total number of neutrons. However, in order to compare the results obtained with different basis sizes, it is useful to define an approximate occupation density, such as:

\begin{equation}
v^{2}_{{\rm cont.}}(\eps)\approx \frac{v^{2}_{\rm disc.}}{\Delta\eps} ,
\end{equation}
where we can approximate this $\Delta\eps$ for the $n$-th pseudoestate as:

\begin{equation}
\Delta\eps \approx \frac{\eps_{n+1}-\eps_{n-1}}{2} .
\end{equation}
Following this prescription, for N$\rightarrow\infty$, the integral of $v^{2}_{\rm disc.}/\Delta\eps$ in Fig.~\ref{occ} will be precisely the total occupation in the continuum.

\begin{figure}
\hspace{0.15\textwidth}
\includegraphics[width=0.5\textwidth]{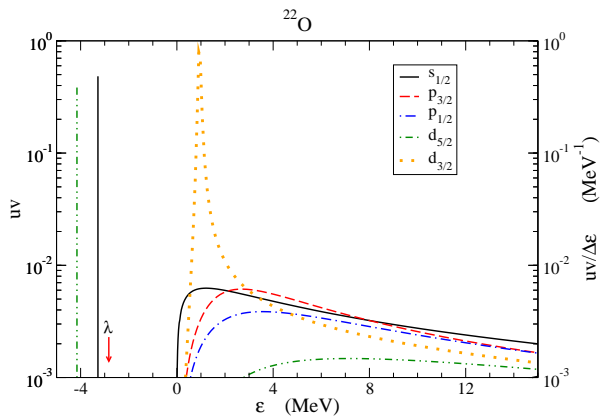}
\caption{\label{uv} (Color online) Product of $uv$ for $^{22}$O for the different partial waves. For the pseudostates this quantity is divided by $\Delta\eps$ so that it is represented in MeV$^{-1}$ (vertical scale on the right). The red arrow indicates the Fermi level.}
\end{figure}

Other interesting observable is the product $uv$ since it is connected with the pair transition density and it should be maximum around the Fermi level. Therefore, this quantity should be more sensitive to the presence of the continuum for weakly bound nuclei. We show the results in Fig.~\ref{uv} for $^{22}$O. For the bound states we represent the direct product $uv$, whereas for continuum states we again divide by $\Delta\eps$.

\subsubsection{\label{occup} Dependence with the binding of the system}


In this section we will concentrate on the $^{22}$O nucleus and focus on the sensitivity of the occupations to the binding of the system. In order to do so, we will reduce the strength of the central potential that determines the single-particle structure. Thus, last bound levels: 1$d_{5/2}$ and 2$s_{1/2}$ will become progressively more weakly bound. Moreover, the Fermi level will get closer to zero.



\begin{figure}
\hspace{0.15\textwidth}
\includegraphics[width=0.5\textwidth]{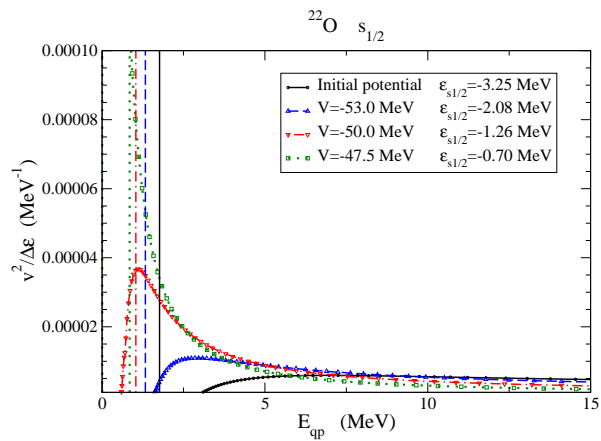}
\caption{\label{swaves} (Color online) Density of occupation in the continuum for $^{22}$O for the $s$-waves when varying the depth of the potential $V$. The vertical lines show the position of the bound state.}
\end{figure}

We start by looking at the occupation of the $s$-waves. Results are shown in linear scale in  Fig.~\ref{swaves}. The x-axis represents the quasi-particle energy:

\begin{equation}
E_{qp}=\sqrt{(\eps-\lambda)^2+\Delta^2} ,
\end{equation}
so that the starting point for continuum states depends on the Fermi level $\lambda$. We see in Fig.~\ref{swaves} how the maximum in the occupation distribution moves to smaller quasi-particle energies when $\lambda\rightarrow0$. In addition, the coupling with a low-lying $s$-state increases the occupation as already reported in~\cite{Ham04} within a different framework. Its presence is only appreciable for very small binding energies. The green dotted line in Fig.~\ref{swaves} corresponds to $\lambda=-0.00004$~MeV.

\begin{figure}
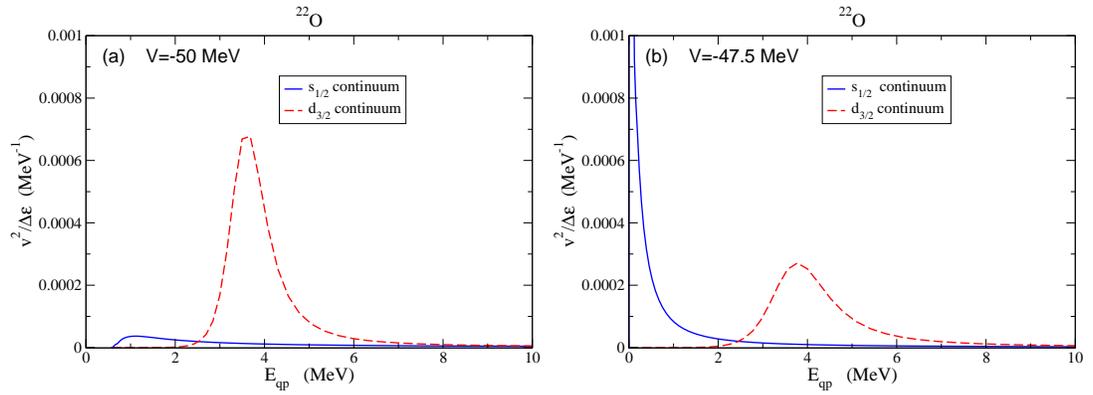

\hspace{0.15\textwidth}\includegraphics[width=0.45\textwidth]{noresvsres_3a.eps}\hspace{0.01\textwidth}\includegraphics[width=0.45\textwidth]{noresvsres_3b.eps}
\caption{\label{noresvsres} (Color online) Density of occupation in the $s_{1/2}$ (solid lines) and $d_{3/2}$ (dashed lines) continua for $^{22}$O as a function of the quasi-particle energy for the two smallest depths of the potential: $V=-50$~MeV in the left pannel and$V=-47.5$~MeV in the right pannel. 
}
\end{figure}

The difference with respect to~\cite{Ham04} is that here we do not include only $s$-waves, but also a $d_{5/2}$ bound state and $s$, $p$, and $d$ continuum, where a $d_{3/2}$ resonance appears. We compare the strengths of the resonance and of the non-resonant continuum for this extreme situation in Fig.~\ref{noresvsres} for the last two $V$ values considered in the previous picture. Again notice that the figure is in linear scale. We might conclude that the influence of the non-resonant continuum becomes important for really weakly-bound systems. However, for this case, we need to go to really extreme situations to overcome the dominance of the resonant part.

\begin{figure}
\hspace{0.15\textwidth}
\includegraphics[width=0.5\textwidth]{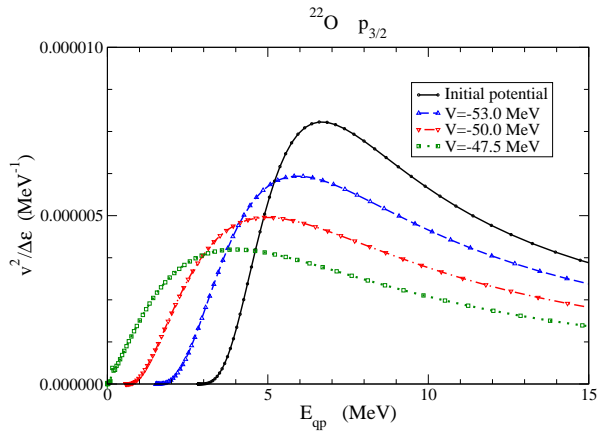}
\caption{\label{pwaves} (Color online) Density of occupation in the $p_{3/2}$ continuum for $^{22}$O as a function of the quasi-particle energy when the depth of the potential $V$ is varied.}
\end{figure}

\begin{figure}
\hspace{0.15\textwidth}
\includegraphics[width=0.5\textwidth]{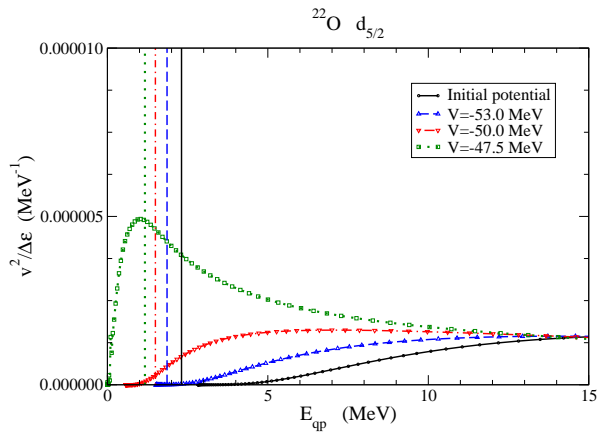}
\caption{\label{dwaves} (Color online) Density of occupation in the $d_{5/2}$ continuum for $^{22}$O as a function of the quasi-particle energy when the depth of the potential $V$ is varied.  The vertical lines show the position of the bound state.}
\end{figure}

The last question is whether $p$-waves follow the same behavior of the $s$-waves. This can give us a clue of the role of the centrifugal barrier. We show in Fig.~\ref{pwaves} the occupations for the $p_{3/2}$-waves. Here, the density of occupation does not change significantly. However, if we look at the $d_{5/2}$ continuum in Fig.~\ref{dwaves}, the occupation for the less bound potential used here show an increase. Therefore, we might conclude that the enhancement is related with the presence of a low-lying state with same angular momentum.

\begin{figure}
\hspace{0.15\textwidth}
\includegraphics[width=0.5\textwidth]{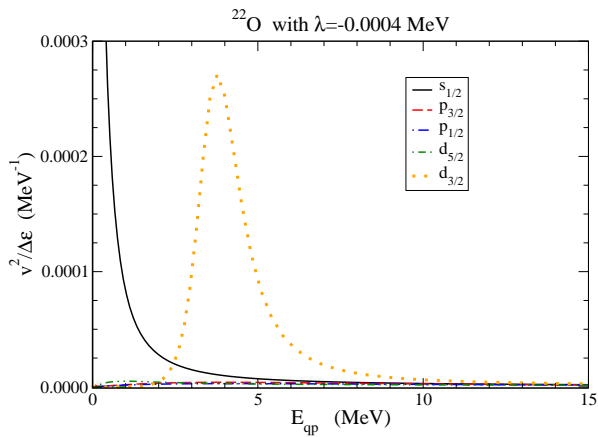}
\caption{\label{occ_really_wb} (Color online) Density of occupation in the continuum for $^{22}$O for the different partial waves for a single-particle potential with central strength $V=-47.5$~MeV.  }
\end{figure}

Again, this occurs only for the extreme situation with an almost zero Fermi level. We show in Fig.~\ref{occ_really_wb} the distribution of the occupation for the continuum including all partial waves considered in the calculation. The total occupation for the main partial waves is 0.4048 (6.75\%) for the $s_{1/2}$, 0.0002 (less than 0.01\%) for the $d_{5/2}$, and 0.0023 (0.04\%) for the $d_{3/2}$.


More examples of occupation in the continuum can be seen in~\cite{Zha11,Gra01,Pas13}. In~\cite{Gra01} neutron rich nickel isotopes are studied, finding an occupation distribution for the $s$-waves for $^{84}$Ni similar to the $V=-50$~MeV case. In~\cite{Zha11,Pas13} mainly occupation for resonances is studied.

\subsection{Carbon isotopes}

For carbon isotopes we start from $^{12}$C as a core and we keep the same single-particle structure as in~\cite{IdB12,IdB12b}. We keep, therefore, the same potential, a Woods-Saxon with a spin-orbit term whose strengths were fitted to reproduce the bound states in $^{13}$C and the $d_{3/2}$ resonance at $2.2$~MeV. It produces as well a broad $f_{7/2}$ resonance around $10$~MeV. In order to see also the effect of this resonance, we include partial waves up to $\ell=3$.

For the pairing strength, we include a volume term $V_{I}$ with $\eta=1$ such that it reproduces the same binding energy for $^{14}$C found in~\cite{IdB12,IdB12b}. The geometry of the density is the same as in the single-particle potential. The strength of $V_{I}$ is later renormalized for the following even carbon isotopes so that it follows the known $1/A$ dependence. The energy cut-off is set at 20~MeV.

\begin{figure}
\hspace{0.15\textwidth}
\includegraphics[width=0.5\textwidth]{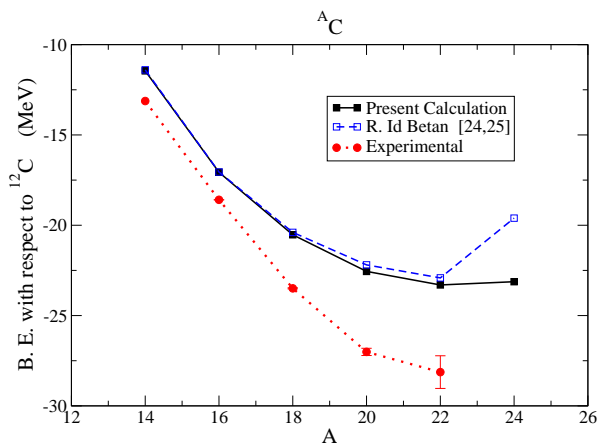}
\caption{\label{becarbon} (Color online) Binding energies for the even neutron rich isotopes of carbon. The experimental energy of $^{12}$C has been substracted in order to compare the experimental data (error bars are sometimes smaller than the point size) to the theoretical calculations with $^{12}$C as a core. }
\end{figure}

\begin{figure}
\hspace{0.15\textwidth}
\includegraphics[width=0.5\textwidth]{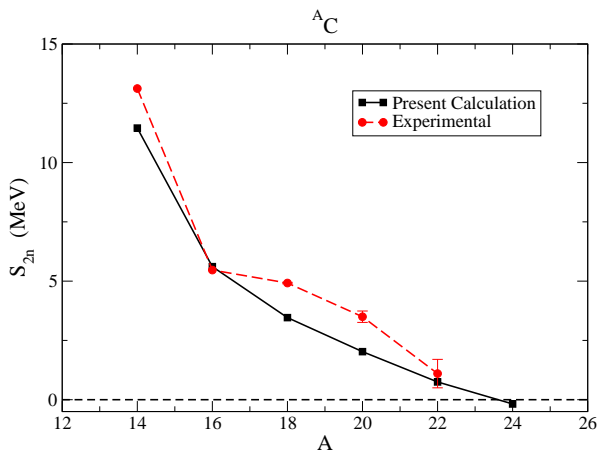}
\caption{\label{s2ncarbon} (Color online) Two-neutron separation energies for the even neutron rich isotopes of carbon.}
\end{figure}

In~\cite{IdB12,IdB12b}, the single-particle continuum is included via the CSPLD. The CSPLD renormalizes the pairing strength and also prevents according to~\cite{IdB12,IdB12b} occupations larger than the expected degeneracy in the continuum. PS have been shown to carry the information about the single-particle level density~\cite{Lay10,Lay12} which can be reobtained from the overlap of the PS with the actual scattering wave-functions. As seen in section~\ref{occup}, it is not compulsory to impose externally a level density. The final occupancies in the continuum, using only the bare PS, follow the behavior shown in~\cite{Ham04,Gra01,Pas13}. Moreover, the $d_{3/2}$ resonance never carries more than 4, i.e. $2j+1$, neutrons.

In Fig.~\ref{becarbon}, we compare the binding energies obtained for neutron rich carbon isotopes within the present work with those from~\cite{IdB12,IdB12b}. In~\cite{IdB12}, the CSPLD is included in the formalism as an attempt to solve the problems of BCS with the Fermi gas~\cite{Dob96}. In~\cite{IdB12b} Richardson equations are solved which should give an exact solution going beyond the BCS approximation. A good agreement is found between the results in~\cite{IdB12b} and ours. Also in both calculations $^{24}$C is found to be unbound. Comparison with the experimental data is also included. In Fig.~\ref{s2ncarbon} we show the comparison with the experimental data for the two-neutron separation energies. 

\begin{figure}
\hspace{0.15\textwidth}
\includegraphics[width=0.5\textwidth]{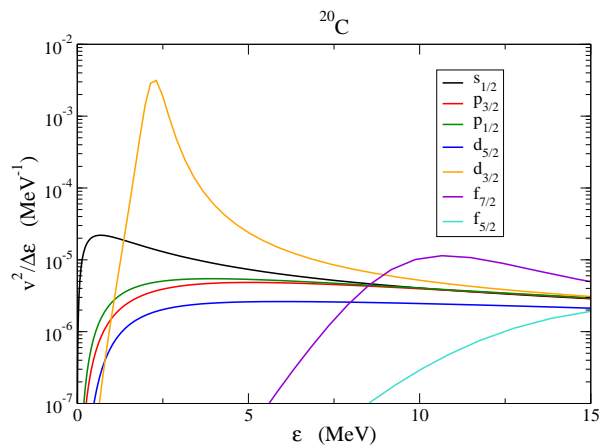}
\caption{\label{occ20C} (Color online) Density of occupation in the continuum for $^{20}$C for the different partial waves.}
\end{figure}

\begin{figure}
\hspace{0.15\textwidth}
\includegraphics[width=0.5\textwidth]{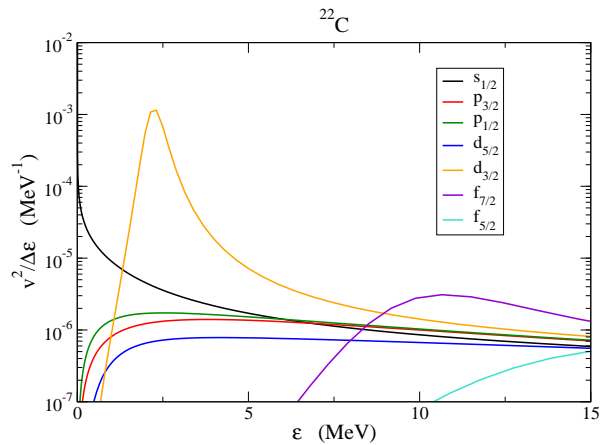}
\caption{\label{occCarbon} (Color online) Density of occupation in the continuum for $^{22}$C for the different partial waves.}
\end{figure}

The occupation distributions for $^{20}$C and $^{22}$C are shown in Fig.~\ref{occ20C} and Fig.~\ref{occCarbon} respectively. Compared with $^{22}$O we see immediately that these carbon isotopes are less bound as expected from the smaller number of protons. The 2$s_{1/2}$ state at $-1.85$~MeV together with the Fermi level close to the threshold  make so that the occupations of the $s$-wave look like the last two cases shown for $^{22}$O, i.e. for $V=-50$~MeV and $V=-47.5$~MeV. We also see the contribution from the two resonances: the narrow $d_{3/2}$ and the broad $f_{7/2}$ resonance. No occupations are shown for this case in~\cite{IdB12,IdB12b}.

As an additional comment, we see here that the occupation of the resonance is carried by more than one pseudostate. For a small size of the basis, one may have only one pseudostate carrying the behavior of the resonance, so called resonant pseudostate. However, when the size is large, there are situations where the resonance is split into several pseudostates as here. This fact has to be taken in consideration when using pseudostates for BCS including only the resonant part. The selection of the pseudoestate for representing the resonance should fit the stabilization method, see Ref.~\cite{Haz70,Lay12proc,Mor12a}. As already mentioned, the main advantage of the discretization with pseudostates is the natural treatment of the continuum where resonant and non-resonant continuum appears on the same footing.

\begin{figure}
\hspace{0.15\textwidth}\includegraphics[width=\textwidth]{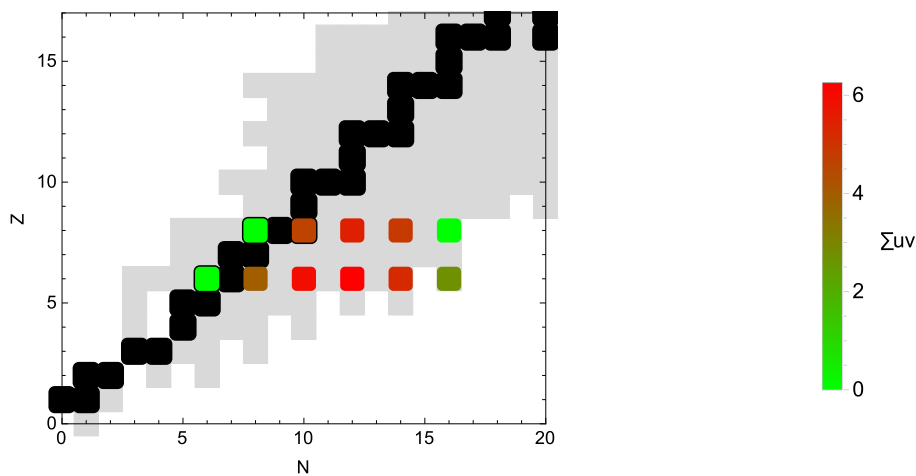}
\caption{\label{coll} (Color online) Segr\`e chart with the total sum of the product $uv$ for the different Oxygen and Carbon isotopes. Stable nuclei are shown in black and those nuclei which are stable with respect to particle emission, in light grey.}
\end{figure}

To conclude this results section, we compare both isotopic chains in Fig.~\ref{coll} by means of total $uv$ product. As we have already mentioned, this quantity is related to the pair transition density reflecting the collectivity of the nucleus. We note in both chains a maximum around $N=12$. For $N=16$ there is significant difference between both chains, $^{24}$O behaves like a closed sub-shell nucleus whereas in $^{22}$C the product of $uv$ is still far from zero. This quantity is zero by construction for $^{16}$O and $^{12}$C.





\section{Conclusions}

We have demonstrated that the THO basis can be profitably used in the solution of the BCS equations with a discretized single-particle continuum. 
The single-particle Hamiltonian is constructed with a Wood-Saxon potential with parameters adjusted to just one isotope for each chain in order to reproduce main bound states and first resonances. This Hamiltonian is discretized 
in the THO basis. Once discretized, Generalized BCS equations can be solved following the traditional formulation for bound states. The different pairing couplings are calculated with a density-dependent delta interaction (DDDI).

The Locale Scale Transformation (LST) applied to the HO wavefunctions allows us to distribute the discretized states along the single-particle continuum almost on demand. This characteristic makes the procedure explained here an easy and practical tool in the search for threshold effects related to pairing within weakly bound systems. Such malleability is not present in the bases commonly used for the same scope.

This method has been applied to Oxygen and Carbon isotopic chains. For the oxygen case we use the potential from Ref.~\cite{Spa00} that reproduces the low-lying spectrum of $^{17}$O. The DDDI is fixed to reproduced the two-neutron separation energy in $^{18}$O and later renormalized for each even isotope according the well known factor $1/A$. With this set of parameters, the overall trend of the binding energies is reproduced and the neutron drip-line appears in the right place. However, a small effect of the continuum is found, as expected regarding the large separation between the last bound state and the continuum threshold.

In order to investigate how the system and the importance of the continuum evolves when this continuum threshold is reduced, we focused on $^{22}$O and reduced the binding potential. An increase of the occupation near the threshold is found for $s-$ and $d_{5/2}-$waves whereas a decrease is found for $p-$waves. We understand this is due to the presence of a low-lying bound state with same spin and parity. The pairing of the low-lying continuum to this last bound state is strong and increase as long as the Fermi level approaches the continuum threshold. This effect was previously found in more complex calculation for different cases, but only for $s-$states~\cite{Ham04,Gra01}.

Moreover, when the separation between the Fermi level and the continuum threshold goes to zero, the contribution of the $d_{3/2}$ resonance to the total occupation is reduced in favor of these threshold contributions. This opens the possibility of situations where the presence of the low-lying continuum is more important than the resonances in the ground state of a many-body system. However, it seems to happen only for extreamely weakly bound situations. This is the most interesting contributions of the present manuscript. This conclusion is also based in the fact that the present model is able to treat resonant and non-resonant continuum on the same footage.

Same procedure has been applied to Carbon isotopes, fitting the parameters following a similar work by Id Betan~\cite{IdB12,IdB12b} in order to compare both methods. Our results for the binding energies of the different isotopes coincide with this former work. The drip-line is also found at the same place. However, experimental binding energies are larger. As already stated in~\cite{IdB12,IdB12b}, this should be attributed to the deformation.

The distributions of the occupation in $^{20}$C and $^{22}$C show similar features to those found for $^{22}$O when reducing the binding energy of the single-particle states. In  $^{22}$C, the importance of the $s$ continuum is almost of the same order than the $d_{3/2}$ resonance. However, deformation can change this situation, so that a BCS description based on Nilsson rather than spherical states might be more accurate.

We are aware of the fact that, in coordinate space, densities obtained with a continuum formulation of the BCS show certain drawbacks which could be connected with the presence of an unphysical gas of neutrons. However, the present formalism satisfies a compromise between accuracy and simplicity needed to pinpoint, control and separate the different ingredients governing pairing in weakly-bound systems.

In conclusion, the simplicity of the present model has given us the flexibility required for exploring the evolution of many-body systems when becoming weakly bound. We found situations where the non-resonant continuum cannot be neglected in favor of resonances as previously commonly stated. However, these facts should be explored increasing the complexity step by step in order to have a clear view of the behavior of weakly bound system. The ability of the THO basis to concentrate states at the really low-energy continuum, reducing the overall computational demand, might be of great help in future attempts.

\ack
This work has been supported by MIUR research fund PRIN 2009TWL3MX, by the Spanish Ministerio de Econom\'{i}a y Competitividad under Project FIS2014-53448-C2-1-P by Junta de Andaluc\'{i}a under group number FQM-160 and Project P11-FQM-7632, and by 
the Consolider-Ingenio 2010 Program CPAN (CSD2007-00042). The research leading to these results has received funding from the European Commission, Seventh Framework Programme (FP7/2007-2013) under Grant Agreement nº 600376. J.A.L. is a Marie Curie Piscopia fellow at the University of Padova.

\bibliographystyle{iopart-num}
\bibliography{bcs_wbs_jpg}

\end{document}